\documentstyle[12pt]{article}
\newcommand\be{\begin{equation}}
\newcommand\ee{\end{equation}}
\newcommand\bea{\begin{eqnarray}}
\newcommand\eea{\end{eqnarray}}

\newcommand{\fatalpha}{{\bf \alpha \kern -0.44em \alpha}}
\newcommand{\fatsigma}{{\bf \sigma \kern -0.54em \sigma}}
\newcommand{\tpchi}{{\bf \chi \kern -0.35em \chi}}
\newcommand{\llambda}{{\bf \lambda \kern -0.45em \lambda}}

\bibliography{plain}
\title{\bf Hierarchy of rational order families of chaotic maps with an invariant measure}
\vspace{20mm} \author{ M. A.
Jafarizadeh$^{a,b,c}$\thanks{E-mail:jafarizadeh@tabrizu.ac.ir}$\;$
, M. Foroutan$^{b,c}$\thanks{E-mail:m.foroutan@tabrizu.ac.ir}$\;$
and S. Ahadpour$^{a,b,c}$\thanks{E-mail:s.ahadpour@tabrizu.ac.ir}
.\\
\\ $^a${\small Department of Theoretical Physics and Astrophysics, Tabriz
University, Tabriz 51664, Iran.}
\\ $^b${\small Institute for Studies in Theoretical Physics and Mathematics,
Teheran 19395-1795, Iran.}
\\ $^c${\small Excellency of Physics, Physics Department, Tabriz University, Tabriz 51664,
Iran.}}
\newpage
\begin{document} \maketitle \vspace{15mm}
\newpage
\begin{abstract}
We introduce an interesting hierarchy of rational order chaotic
maps that posses an invariant measure. In contrast to the
previously introduced hierarchy of chaotic maps
\cite{J1,J2,J3,J4,J5}, with merely entropy production, the
rational order chaotic maps can simultaneously produce and consume
entropy . We compute the Kolmogorov-Sinai entropy of theses maps
analytically and also their  Lyapunov exponent numerically, where
that obtained numerical results support the analytical
calculations.
 {\bf PACs
numbers:05.45.Ra, 05.45.Jn, 05.45.Tp }
\end{abstract}
\pagebreak \vspace{70mm}
\section{Introduction}
There has been some attempts \cite{J1,J2,J3,J4,J5,Go} at
introducing the hierarchy of chaotic maps with an invariant
measure in the recent years. The objective of these papers is to
describe the dynamic behavior of chaotic maps using
Kolmogorov-Sinai entropy. These hierarchies of chaotic maps are
of interest as models for describing of behavior of dynamical
systems. As an example, the random chaotic maps have attracted
the attention of physicists as models of convection by
temporarily irregular fluid flows \cite{yu}. Once a map is
determined, the long term statistical behavior is described by a
probability density function, which can be obtained either by
solving the Frobenius-Perron equation \cite{Boy} or can be
estimated by measurement of the system. Therefore, the complexity,
non-linearity and non-stationarity of physical, chemical,
biological, physiological and financial systems
\cite{Cl,Va,Mi,Vi,Iv,We} have been of main interest in introducing
the new hierarchy  of the chaotic maps. On the other hand, the
sensitively to the initial condition, control parameter and
ergodicity which have tight relationships with the requirement of
pseudo-random coding and cryptography \cite{Bro,Fri} are examples
of interesting features of chaotic systems and it is natural idea
to use chaos as a new source to construct new encryption systems
\cite{Shu}. \\In present paper, we introduce the rational order
families of chaotic maps as a new hierarchy of chaotic map with an
invariant measure. The Kolmogorov-Sinai entropy of these chaotic
maps can be calculated analytically by using their invariant
measure. An interesting property of these  chaotic maps is their
ability in simultaneous production and consumption of entropy.
Additionally, being a measurable dynamically systems, so it can
be studied analytically.\\ The paper is organized as follows: In
Section 2, we introduce the rational order families of chaotic
maps. In Section 3, the invariant measure of these maps are given
and in Section 4, we review the Kolmogorov-Sinai entropy and
compute it for the rational order chaotic maps. Finally in
Section 5 we calculate the Lyapunov exponent numerically and
compare the results of simulation with analytically calculated
Kolmogorov-Sinai entropy. The last sections contains our
conclusion and two appendices. In these appendices we have
calculated the invariant measure of the rational order families
of chaotic maps via two different methods.
\section{Hierarchy of rational order
families of chaotic maps with an invariant
measure}\setcounter{equation}{0} We first review hierarchy of
one-parameter chaotic maps which can be used in the  construction
of families of rational order chaotic maps with an invariant
measure. The one-parameter chaotic maps \cite{J1} are defined as
the ratio of polynomials of degree $\mathbf{N}$:
$$\phi_{N}^{1}(x,\mbox{a})=\frac{\mbox{a}^{2}(1+(-1)^{N}\quad
_{2}F_{1}(-N,N,\frac{1}{2},x))}{(\mbox{a}
^{2}+1)+(\mbox{a}^{2}-1)(-1)^{N}\quad_{2}F_{1}(-N,N,\frac{1}{2},x)}$$$$=\frac{\mbox{a}^{2}(T_{N}(\sqrt{x}))^{2}}{1+(\mbox{a}^{2}-1)(T_{N}(\sqrt{x})^{2})
}$$
$$\phi_{N}^{2}(x,\mbox{a})=\frac{\mbox{a}^{2}(1-(-1)^{N}\quad_{2}F_{1}(-N,N,\frac{1}{2},(1-x)))}
{(\mbox{a}^{2}+1)-(\mbox{a}^{2}-1)(-1)^{N}
\quad_{2}F_{1}(-N,N,\frac{1}{2},(1-x))}$$
$$=\frac{\mbox{a}^{2}(U_{N}(\sqrt{(1-x)}))^{2}}{1+(\mbox{a}^{2}-1)(U_{N}(\sqrt{(1-x)})^{2})}$$
where $\mathbf{N}$ is an integer greater than one. Also,
$$_{2}F_{1}(-N,N,\frac{1}{2},x)=(-1)^{N}\cos (2N \arccos
\sqrt{x})=(-1)^{N} T_{2N}(\sqrt{x})$$ is the hypergeometric
polynomials of degree $N$ and $T_{N}(U_{n}(x))$ are  Chebyshev
polynomials of type I (type II), respectively. Here in this paper
we are concerned with their conjugate maps which are defined as:
\begin{equation}
\left\{\begin{array}{l}\tilde{\phi}_{N}^{(1)}(x,\mbox{a})=h\circ
\phi_{N}^{(1)}(x,\mbox{a})\circ
h^{-1}=\frac{1}{\mbox{a} ^{2}}\tan ^{2}(N \arctan \sqrt{x}),\\
\tilde{\phi}_{N}^{(2)}(x,\mbox{a})=h\circ
\phi_{N}^{(2)}(x,\mbox{a})\circ h^{-1}=\frac{1}{\mbox{a}
^{2}}\cot ^{2}(N \arctan \frac{1}{\sqrt{x}}).
\end{array}\right.
\end{equation}
Conjugacy means that invertible map $h(x)=\frac{1-x}{x}$ maps
$I=[0,1]$ into $[0,\infty)$.
\\Now, in order to generalize the above hierarchy of integer order chaotic
maps to the hierarchy of rational order chaotic maps with an
interesting property of simultaneous production and consumption
of entropy,  we need to replace $x_{n+1}$ with a non-linear
function of $x_{n+1}$, particularly a non-linear function of
chaotic maps of above type. But in order  to have a single-valued
map, we will take one of inverse branches of above nonlinear
functions for $x_{n+1}$ in each step with a probability equal to
the  probabilities of occurrence of the branches in iteration of
the maps. As we will show in section 3, the probability of
occurrence of each branch is equal to the integral of invariant
measure of the map over the corresponding domain of the same
branch. Therefore we can define these maps as:
\begin{equation}
 x_{n+1,k}=g_{2,k}^{ -1}\circ g_1(x_n),\;\;i,j\in \{1,2 \})\;\;\mbox{with
probability }\; P_{k},
\end{equation}
where functions $g_1$ and $g_2$ can be chosen as  one the
functions given in (2-1)  and $P_{k}$ are probabilities of
occurrence of  inverse branches $ g_{2,k}^{
 -1}$ of the map $ g_2$ in  its iteration. As  we will see at the end of this
section, the existence of an invariant measure
 will impose a relation between their parameters.\\
Off course the  functions given in (2-1) is not the only choice
for the functions $g_1$ and $g_2$ that leads to the hierarchy of
rational order chaotic maps with an invariant measure. Obviously
the following choices of  the functions $g_1$ and $g_2$
 \be
\begin{array}{ccccc} \textbf{a)} & \frac{1}{\mbox{a}^2}\tan^{2}(N arccot\sqrt{x}),& &
\textbf{b)} & \frac{1}{\mbox{a}^2}\cot^{2}(N arctan \sqrt{x}), \\
\textbf{c)} & \frac{1}{\mbox{a}^2}\cot^{2}(N arccot \sqrt{x}),&  & \textbf{d)} & \frac{1}{\mbox{a}}|\tan(N arctan |x|)|, \\
\textbf{e)} &\frac{1}{\mbox{a}}|\tan(N arccot |x|)|,& &   \textbf{f)} & \frac{1}{\mbox{a}}|\cot(N arctan |x|)|, \\
\textbf{g)} & \frac{1}{\mbox{a}}|\cot(N arccot |x|)|,&  & \textbf{h)} & \frac{1}{\mbox{a}}\tan(N arctan x), \\
\textbf{I)} & \frac{1}{\mbox{a}}\tan(N arccot x), & &  \textbf{J)} &\frac{1}{\mbox{a}} \cot(N arctan x), \\
\textbf{k)} & \frac{1}{\mbox{a}}\cot(N arccot x)&
&\nonumber\end{array}\ee lead to  the hierarchy of rational order
chaotic maps of trigonometric types (with an invariant measure),
where some of them are equivalent to each others up to conjugacy.
Also with the choices of $g_1$ and $g_2$ as \cite{J4} \be
\begin{array}{ccccc} \textbf{a)} & \frac{1}{\mbox{a} ^{2}}{\bf sc}^{2}(N {\bf
sc}^{-1}( \sqrt{x}))& &\textbf{b)}, & \frac{1}{\mbox{a} ^{2}}{\bf
cs}^{2}(N {\bf cs}^{-1}(\sqrt{x}))\end{array},\ee
 we get the
Hierarchy of elliptic rational order chaotic maps of ${\bf cs}$
and ${\bf sc}$ types, where their invariant measure can be
obtained  for  small enough values of module k of elliptic
functions. Also it is possible to choose the function $g_1$ and
$g_2$ as one of the combined chaotic maps of Ref. \cite{J3}.\\
Here in this paper we will consider the  hierarchy of rational
order maps with
\begin{equation}g_1(a_1,N_1,x_n)=\frac{1}{\mbox{a}_1}\tan(N_{1}\arctan
x_{n})\; \mbox{and}\;
g_2(a_2,N_2,x_{n+1})=\frac{1}{\mbox{a}_2}\tan(N_{2}\arctan
x_{n+1}),\end{equation} i.e., we have
\begin{equation}
x_{n+1,k_2}=\tan\left(\frac{\arctan(\frac{\mbox{a}_2}{\mbox{a}_1}\tan(N_{1}\arctan
x_{n}))}{N_2}+\frac{k_2 \pi }{N_2}\right)\;\mbox{with
probability}\; P_{k_2},\;k_2=1,2,...,N_2,
\end{equation}
where $N_{1}$ and $N_{2}$  are integer greater than one and
$\mbox{a}_{1}$ and $\mbox{a}_{2}$ are control parameters. As we
are going to see in section 3, the maps (2-6) posses an invariant
measure provided that we choose the parameters $a_{1}$ and
$a_{2}$ in the form  given in Equations (3-12) and (3-13),
respectively. As an example we consider the following map for
$N_1=3$ and $N_2=2$:

\begin{equation}
x_{n+1,\pm}=\frac{ \mbox{a}_{1}}{
\mbox{a}_{2}}\times\frac{1-3x_{n}^{2}}{3x_{n}-x_{n}^{3}}\pm
\sqrt{1+(\frac{ \mbox{a}_{1}}{
\mbox{a}_{2}}\times\frac{1-3x_{n}^{2}}{3x_{n}-x_{n}^{3}})^2}\;\;\mbox{with
probabilities}\;\; P_{\pm}=\frac{1}{2}.
\end{equation}

\section{Invariant measure}\setcounter{equation}{0}
 A dynamical system even time-discrete
one-variable system has a number of possible types of behavior.
The system can be in a fixed point and nothing changes, the
trajectory of the system may also be on a cycle with a certain
period. Fixed point and periodic orbits may be stable or
unstable.We are usually interested in an invariant measure $ \mu
$, i.e. a probability measure that does not
 change under the dynamics.
The probability measure $\mu$ on $[0,1]$ is an Sinai-Rulle-Bowen
(SRB)  measure as an invariant measure which describes
statistically stationary states of system and
absolutely continues with respect to Lebesgue measure.\\
Now in order to determine the invariant measure of the analytical
system described by the maps given in  (2-6), we can write it as
combination of the maps $g_1$ and $g_{2,k_2}^{-1}$ (as the
$k_2$-the inverse branch of $g_{2}$) in the following form:
\begin{equation}
x_{n+1,k_2}=g_{2,k_2}^{-1}\circ g_{1}(x_n)\quad\quad\mbox{with
probability}\; P_{k_2}\quad\quad k_2=1,...,N_{2}\end{equation}
with $g_1$ and $g_2$ given in (2-5).

Obviously the function $g_2(.,\mbox{a}_{2},N_2)$ maps, its $N_2$
inverse  branches $x_{n+1,k_2}\; k_2=1,2,...,N_2$ with
corresponding different domains $\Delta x_{n+1,k_2}\;(\Delta
x_{n+1,i}\bigcap \Delta x_{n+1,j}=\emptyset$ for $i\neq
j=1,2,...,N_2$) into the same region. Therefore, if denote its
value by $y$ for different values of its argument then the map
(3-1) can be written as:
\begin{equation}
g_{2}(x_{n+1},\mbox{a}_{2},N_2) =y= g_{1}(x_{n},\mbox{a}_{1},
N_1),\end{equation} irrespective of to which branch or domain, the
output $x_{n+1}$ belongs (see Fig. 1). But in order to have a
single output or single valued dynamical map, we have to consider
only one of possible $x_{n+1}$ in each step with some
probabilities or weights. Certainly the most natural weight of a
given branch  is the corresponding probability of its occurrence
in infinite iteration of map $y=g_2(.,\mbox{a}_{2},N_2)$, where
it can written in terms of its invariant measure $\mu_{g_2}$ as,
\begin{equation}
\mbox{P}(\mbox{occurrence of}\; k_2-\mbox{the
branch})=\int_{\Delta x_{n+1,k_2}}\mu_{g_2}(x)dx.
\end{equation}
 Therefore the invariant measure
of this map should satisfy the following  Frobenius-Perron
integral equations:
\begin{equation}
\mu(y)=\int_{0}^{1}\delta(y-g_{1}(x_{n},\mbox{a}_{1},N_1))\mu(x_{n})dx_{n}
\end{equation}
and
\begin{equation}
\mu(y)=\int_{0}^{1}\delta(y-g_{2}(x_{n+1},\mbox{a}_{2}))\mu(x_{n+1})dx_{n+1},
\end{equation}
which are equivalent to:
\begin{equation}
\mu(y)=\sum_{x_{n,k_{1}}\in
g_{1}^{-1}(y)}\mu(x_{n,k_{1}})|\frac{dx_{n,k_{1}}}{dy}|
\end{equation}
and
\begin{equation}
\mu(y)=\sum_{x_{n+1,k_{2}}\in
g_{2}^{-1}(y)}\mu(x_{n+1,k_{2}})|\frac{dx_{n+1,k_{2}}}{dy}|,
\end{equation}
where
$$x_{n,k_{1}}=\tan (\frac{1}{N_{1}} \arctan (\mbox{a}_1y)+\frac{k_{1}\pi}{N_{1}}),\quad\quad k_{2}=1,...,N_{1}.$$and
$$x_{n+1,k_{2}}=\tan (\frac{1}{N_{2}} \arctan (\mbox{a}_2y)+\frac{k_{2}\pi}{N_{2}}),\quad\quad\quad k_{2}=1,...,N_{2}.$$
The invariant measure $\mu(y)$ for $g_{i}(x)$ can be written as:
\begin{equation}
\mu_{g_{i}(x)}(y)=\sum_{k_{i}=1}^{N_{i}}
\frac{\mbox{a}_{i}}{N_{i}}\left(\frac{1+x_{n,ki}^{2}}{1+(\mbox{a}_{i}y)^{2}}\right)\mu_{g_{i}}(x),
\quad\quad\quad i=1,2.
\end{equation}
Assuming that $\mu(x)$ has the following form:\begin{equation}
\mu(x)=\frac{\sqrt{\beta}}{\pi(1+\beta x^{2})}
\end{equation}
where for $\beta=1$, it reduces to the invariant measure which has
already applied to pushout measure \cite{Dou}, expression (3-6)
reduces to
\begin{equation}
 \frac{1+(\mbox{a}_{i}y)^{2}}{1+\beta
y^{2}}=\sum_{k_i=1}^{N_{i}}\frac{\mbox{a}_{i}}{N_{i}}(\frac{1+x_{n,ki}^{2}}{1+\beta
x_{n,ki}^{2}}), \quad\quad\quad i=1,2.
\end{equation}
By comparing  of both sides of Equation (3-8), we can determine
$\mbox{a}_{i}$, ($i=1,2$) as
\begin{equation}
\mbox{a}_{i}=\frac{\sum_{k=0}^{[\frac{N_{i}}{2}]}C_{2k}^{N_{i}}\beta^{k}}{\sum_{k=0}^{[\frac{N_{i}-1}{2}]}C_{2k+1}^{N_{i}}\beta^{k}},
\end{equation}
 for even values of $N_{i}$, and
\begin{equation}
\mbox{a}_{i}=\frac{\sum_{k=0}^{[\frac{N_{i}-1}{2}]}C_{2k+1}^{N_{i}}\beta^{k}}{\sum_{k=0}^{[\frac{N_{i}}{2}]}C_{2k}^{N_{i}}\beta^{k}},
\end{equation}
 for odd values of $N_{i},$(for proof see Appendix A). \\Therefore $\mbox{a}_{1}$ and $\mbox{a}_{2}$ depend on the
parameter $\beta$ and integers $N_{1}$ and $N_{2}$, respectively.
Also to make the paper more
readable, we have derived the invariant measure of the map \\
$y=\frac{1}{4}\tan(4arctan x)$   by using Shure's invariant
polynomials in Appendix B.
\section{Kolmogorov-Sinai entropy }\setcounter{equation}{0}
In this section we review first, the  Shannon entropy and then
talk about Kolmogorov-Sinai entropy (for more details see
\cite{Cha}). Consider dynamical system characterized by a certain
iterative map. Let $B=(B_{i},B_{j},\ldots,B_{n})$ be a
decomposition of the unit interval along $x_{n}$. Now we subdivide
each interval $B_{i}$ into say $ \Lambda $ points, and perform
$\zeta $ iterations on each one of them so we make sure that
transients have died out. Then $ \Lambda$ points by then will
spread to other subintervals. A percentage of them will be perhaps
located within the limits of $B_{j}$ . After transients die out
the common area of $F^{\xi}(B_{i})$ and $B_{j}$, e.g.
$F^{\xi}(B_{i})\bigcap B_{j}$ will be express in a non-normalized
way the number of elements of $B_{i}$ reaching $B_{j}$ after $\xi$
iterations. So in normalized form:
\begin{equation}
W^{\xi}(B_{j}/B_{i})=\frac{\mu(F^{\xi}(B_{i})\bigcap
B_{j})}{\mu(B_{j})},
\end{equation}
here $\mu(.)=\int_{c}\mu(x)dx$, where $\mathbf{c}$ is the
pertinent interval. The entropy of the chosen partition or, the
average amount of information needed to locate the system in state
space is given by the Shannonian entropy;
\begin{equation}
S=-\sum_{i}^{\Lambda}{\mu}(B_{i})\log_{2}\mu(B_{i})bits.
\end{equation}
The $\Lambda$ values $\mu(B_{i})$ may be calculated from the
$W_{ij}$ elements from the $(\Lambda-1)$ equations of the linear
system:
\begin{equation}
\mu(B_{i})=\sum_{j=1}^{\Lambda}\mu(B_{j})W_{ij}
\end{equation}
and the normalization condition:
$$ \sum_{j=1}^{\Lambda}\mu(B_{j})=1, $$ where the transition
probability matrix $W_{ij}$ describes the probability of jumping
in one step (iteration) from the element $B_{i}$ of the partition
to the element $B_{j}$. The average amount of information created
by the linguistic system by per transition per unit time is given
by the Kolmogorov-Sinai entropy for the chosen partition; namely;
\begin{equation}
S_{k}=\sum_{i=1}^{\Lambda}\sum_{j=1}^{\Lambda}\mu(B_{i})W_{ij}\log_{2}W_{ij}\mbox{bits}.
\end{equation}
The macroparameter however, characterizing the degree of
grammatical coherence of the created Markovian chain is the mutual
information or transinformation.
\begin{equation}
I(\xi)=\sum_{i=1}^{\Lambda}\sum_{j=1}^{\Lambda}\mu(F^{\xi}(B_{i})\bigcap
B_{j})\log_{2} \frac{\mu(F^{\xi}(B_{i})\bigcap
B_{j})}{\mu(F^{\xi}(B_{i}){\mu}(B_{j}))}\mbox{bits}.
\end{equation}
It stands for the information stored in a symbol along the
sequence about what is going to emerge $\xi$ iterations (or $\xi$
time units ) later, $I(\xi)$ gives the information transferred
between two symbol $\xi$ steps apart. As the number of the
decomposition of the unit interval goes to infinity, in such a way
that, the size of each intervals $(B_{i})$ goes to zero. The
mutual entropy given in  (4-5) reduces to the well known
Kolmogorov-Sinai (KS) entropy which  is given by:
\begin{equation}
h(\mu,g(x,\mbox{a},N))=\int\mu(x) dx
\ln|\frac{dx_{n+1}}{dx_n}|=\int_{-\infty}^{+\infty}\mu(x) dx
\ln|\frac{d}{dx}g(x,\mbox{a}, N)|
\end{equation}
with $g(x,\mbox{a}, N)=\frac{1}{\mbox{a}} (tan(N \arctan x))$
$h(\mu,g(x,\mbox{a},N))$ can be written as:
\begin{equation}
h(\mu,g(x,\mbox{a}),N)=\int
_{-\infty}^{+\infty}\frac{\sqrt{\beta}}{\pi(1+\beta x^{2})} dx
\ln|\frac{N}{\mbox{a}}\times\frac{1+\mbox{a}^{2}y^{2}}{1+x^{2}}|.
\end{equation}
Following the calculating of Ref.[1], one can show that after a
change of variable $\sqrt{\beta}x=\tan\theta$, and using the
integral of type;
\begin{equation}
\frac{1}{\pi}\int_{0}^{\pi}\ln|a+b\cos\theta|=
\left\{\begin{array}{llll} \ln|\frac{a+\sqrt{a^{2}-b^{2}}}{2}|  & & &  |a|>|b|, \\
\ln|\frac{b}{2}|  &  &  & |a|\leq|b|, \end{array} \right.
\end{equation}
 we get the following expression the KS-entropy:
\begin{equation}
h(\mu,g(x,\mbox{a},N))=\frac{1}{\sqrt{\beta}}ln\left[
\frac{N}{\mbox{a}^{3}}\left(\frac{\sqrt{\beta}(\sum_{k=0}^{[\frac{N_{i}}{2}]}C_{2k}^{N_{i}}x^{k})\mbox{a}+\mbox{a}
(\sum_{k=0}^{[\frac{N_{i}-1}{2}]}C_{2k+1}^{N_{i}}x^{k})}{(\sqrt{\beta}+1)(\sum_{k=0}^{[\frac{N_{i}}{2}]}C_{2k}^{N_{i}}x^{k})}\right)^{2}\right].
\end{equation}
Now, we come to calculate  the KS-entropy of fractional order
maps. Before getting to involved with the details of calculation,
we first  talk about simultaneous production and consumption of
entropy in these  maps. Figure 1, gives us an insight to see how
this is possible. In this figure, $N_{1}$ ( $ N_{2}$) corresponds
to the  number of branches $x_{n+1}$( $x_{n}$) of the map at time
$n+1(n)$. The left hand half of figure 1 shows the contraction of
$N_{1}$ identical branches of $x_{n+1}$ into a single branch,
while the right hand half of the figure shows branch out to $
N_{2}$ branches $x_{n}$,  where they  correspond to increment and
decrease of entropy, respectively. It should be reminded that in
contraction of branches, entropy increase due to loss of
information, while in branch out it decreases due to reception of
 information.
Now in order to  calculate  KS-entropy of fractional order
maps(2-6), we should notice that  the right and left halves of
Figure 1, correspond to the right-hand and left-hand sides of
Equation (3-3)( equivalent of  (2-6)). In other words, there are
$N_{1}$ convergent branch $x_{n+1}$ and $ N_{2}$ divergent branch
$x_{n}$ in the left and right halves of Figure 1. Therefore,
according to Figure 1 there are $N_2$ possible final states
$x_{n+1,k_2}$, $k_2=1,2,...,N-2$ with corresponding weights given
in (3-3), where we should take average over their corresponding
KS-entropy given in (4-6). Hence, the KS-entropy of rational map
(2-6) can be written as
$$h(\mu,\mbox{rational order map})=\sum_{k_2=1}^{N_2}\int_{\triangle x_{n+1,k_2}}\mu(x_{n+1},\mbox{a}_2)dx_{n+1}\int_{-\infty}^{+\infty}\mu(x_{n},\mbox{a}_1)dx_{n}\ln|\frac{dx_{n+1}}{dx_{n}}|$$
$$=\sum_{k_2=1}^{N_2}\int_{\triangle x_{n+1,k_2}}\mu(x_{n+1},\mbox{a}_2)dx_{n+1}\int_{-\infty}^{+\infty}\mu(x_{n},\mbox{a}_1)dx_{n}\ln|\frac{dx_{n+1}}{dy}||\frac{dy}{dx_{n}}|$$$$=
\sum_{k_2=1}^{N_2}\int_{\triangle x
_{n+1,k_2}}\mu(x_{n+1},\mbox{a}_2)dx_{n+1}\int_{-\infty}^{+\infty}\mu(x_{n},\mbox{a}_1)dx_{n}(\ln|\frac{dx_{n+1}}{dy}|+\ln|\frac{dy}{dx_{n}}|)$$
\begin{equation}
=\int_{-\infty}^{+\infty}\mu(x_{n},\mbox{a}_1)dx_{n}\ln|\frac{dy}{dx_{n}}|-
\int_{-\infty}^{+\infty}\mu(x_{n+1},\mbox{a}_2)dx_{n+1}\ln|\frac{dy}{dx_{n+1}}|,
\end{equation}
where in the last line above we have used the following
normalization relation
$$
\int_{-\infty}^{+\infty}\mu(x,\mbox{a}_i)dx=1\quad\quad i=1,2.$$
Now, comparing the last line of (4-10) with (4-6), we get the
following expression for  KS-entropy of fractional order map
(2-6):
\begin{equation}
h(\mbox{rational order map})=
h(\mu,\frac{1}{\mbox{a}_1}\tan(N_{1}\arctan x_{n})
)-h(\mu,\frac{1}{\mbox{a}_2}\tan(N_{2}\arctan x_{n} )
\end{equation}
with $h(\mu,\frac{1}{\mbox{a}_i}\tan(N_{i}\arctan x_{n}) )\;
i=1,2$  given in(4-9). \\Obviously  formula (4-11) implies the
simultaneous production and consumption of the entropy, where the
term with positive sign corresponds to the production of the
entropy while the term with minus sign corresponds to the
consumption of the entropy, respectively. Also, it is interesting
to note that maximum value of the entropy  is equal to
$\ln_{2}\frac{N_{2}}{N_{1}}$
that corresponds to, $\mbox{a}_{1}=\mbox{a}_{2}=1$ ( actually this is the main reason for naming these maps as  rational order maps).\\
\section{Lyapunov exponent and simulation:}
\setcounter{equation}{0} A useful numerical way to characterize
chaotic phenomena in dynamic systems is by means of the Lyapunov
exponents that describe the separation rate of systems whose
initial conditions differ by a small perturbation. Suppose that
there is a small change $\delta x(0)$ in the initial state
$x(0)$. At step or time $\mathbf{n}$ this has changed to $\delta
x(n)$ given by:
\begin{equation}
\delta x(n)\approx\delta x(0)|\frac{dx_n}{dx_0}|=\delta
x(0)|\frac{dx_n}{dx_{n-1}}.\frac{dx_{n-1}}{dx_{n-2}}.......\frac{dx_1}{dx_0}|,
\end{equation}
where we have used the chain rule to expand the derivative of
$\frac{dx_n}{dx_0}$. In the limit of infinitesimal perturbations
$\delta x(0)$ and infinite time we get an average exponential
amplification, the Lyapunov exponent $\lambda$,
\begin{equation}
\lambda=lim_{n\rightarrow\infty}\frac{1}{n}\ln|\frac{\delta
x(n)}{\delta
x(0)}|=lim_{n\rightarrow\infty}\frac{1}{n}\ln|\frac{dx_n}{dx_0}|=
lim_{n\rightarrow\infty}\frac{1}{n}\sum_{k=1}^{n}\ln|\frac{dx_k}{dx_{k-1}}|.
\end{equation}
 Similarly the Lyapunov exponent of rational maps (2-6)
can be obtained from formula (5-2) provided that we replace
$x_k$, at random with  $x_{k,\pm}$  at each step $k$ with
probabilities $P_{\pm}=\frac{1}{2}.$  Here in this work we have
simulated Lyapunov exponent of rational order map with $N_{1}=2$
and $N_{2}=3$ for different values of $\beta$, where the result
that obtained supports the analytic calculation of KS-entropy
given in (4-11) for particular case of $N_{1}=2$ and $N_{2}=3$
(see Figure 2).

\section{Conclusion} We
have given a new hierarchy of rational order families of chaotic
maps which presents an interesting description of simultaneous
 production and consumption of  entropy. Using the SRB measure, the
Kolmogorov-Sinai entropy of chaotic maps have been calculated. It
would be interesting to introduce these kinds of maps in higher
dimensions which is under investigation.\nonumber
\section{Appendix A}\setcounter{equation}{0}
Here in this appendix following the prescription of References
\cite{J1,J2,J3,J4,J5}, we prove that \\ the invariant measure
given in  (3-9) satisfies the corresponding PF equations of the
maps\\ $y_i=\frac{1}{\mbox{a}_i}\tan(N_i\arctan x_i),\;i=1,2$,
provided that the parameters $a_1$ and $a_2$ can be expressed in
term of $\beta$ as in formulas (3-11) and (3-12). To do so we we
can write the right hand side of Equation (3-8) as,
\begin{equation}
\frac{\mbox{a}}{N}\sum_{k=1}^{N}\frac{1+x_{k}^{2}}{1+\beta
x_{k}^{2}}=\frac{\mbox{a}}{\beta}+\frac{\mbox{a}(\beta-1)}{N\beta^2}\frac{\partial}{\partial
\beta^{-1}}\ln (\prod_{k=1}^{N}(\beta^{-1}+ x_{k}^{2})),
\end{equation}
where we have omitted the indices $i$ and $n$. Hence, Equation
(3-8) can be written as:
\begin{equation}
 \frac{1+\mbox{a}^{2}y^{2}}{1+\beta
y^{2}}=\frac{\mbox{a}}{\beta}+\frac{\mbox{a}(\beta-1)}{N\beta^2}\frac{\partial}{\partial
\beta^{-1}}\ln (\prod_{k=1}^{N}(\beta^{-1}+ x_{k}^{2})).
\end{equation}
To evaluate the second term in the right hand side of above
formulas we can write the equation
$y=\frac{1}{\mbox{a}}\tan(N\arctan x)$ in the following form:
$$ 0=\mbox{a}y\cos(N\arctan x)-\sin(N\arctan x)
$$ $$
={\frac{1}{(1+x^2)^{\frac{N}{2}}}}\left(\mbox{a}y\sum_{k=0}^{[\frac{N}{2}]}C_{2k}^{N}(-1)^{k}x^{2k}-
x\sum_{k=0}^{[\frac{N-1}{2}]}C_{2k+1}^{N}(-1)^{k}x^{2k}\right),
$$
 $$=\frac{\mbox{constant}}{(1+x^2)^{\frac{N}{2}}}\prod_{k=1}^{N}(x-x_{k})\quad,
$$
where $x_{k}=\tan (\frac{1}{N} \arctan
(\mbox{a}y)+\frac{k\pi}{N})\; k=1,...,N$ are its roots. Therefore,
we have:
$$\frac{\partial}{\partial\beta^{-1}}\ln\left(\prod_{k=1}^{N}(\beta^{-1}+x_{k}^2)\right)=
\frac{\partial}{\partial\beta^{-1}}\left(\ln\left(\prod_{k=1}^{N}(i\sqrt{\beta^{-1}}+x_{k})\right)+\ln\left(\prod_{k=1}^{N}(-i\sqrt{\beta^{-1}}+x_{k})\right)\right)$$
 $$=\frac{\partial}{\partial\beta^{-1}}\ln\left[(1-\beta^{-1})^{\frac{N}{2}}\left(\mbox{a}y\cos(N\arctan (-i\sqrt{\beta^{-1}}))-\sin(N\arctan (-i\sqrt{\beta^{-1}}))\right)\right]$$
$$+\frac{\partial}{\partial\beta^{-1}}\ln\left[(1-\beta^{-1})^{\frac{N}{2}}\left(\mbox{a}y\cos(N\arctan (i\sqrt{\beta^{-1}}))-\sin(N\arctan (i\sqrt{\beta^{-1}}))\right)\right]$$
$$=\frac{\partial}{\partial\beta^{-1}}\ln\left[\mbox{a}^2
y^2(\sum_{k=0}^{[\frac{N}{2}]}C_{2k}^{N}\beta^{-k})^2+\beta^{-1}(\sum_{k=0}^{[\frac{N-1}{2}]}C_{2k+1}^{N}\beta^{-k})^2\right]$$
$$=\frac{\partial}{\partial\beta^{-1}}\ln\left[(1-\beta^{-1})^{N}\left(\mbox{a}^2y^2\cos^2(N\arctan (i\sqrt{\beta^{-1}}))-\sin^2(N\arctan (i\sqrt{\beta^{-1}}))\right)\right]$$
\begin{equation}
=-\frac{N\beta}{\beta-1}+\frac{\beta
N(1+\mbox{a}^2y^2)A(\frac{1}{\beta})B(\frac{1}{\beta})
}{({1-\beta^{-1}})\left((A(\frac{1}{\beta}))^{2}\beta y^2+(B(
\frac{1}{\beta}))^2\right)}\quad,
\end{equation}
with polynomials $ A(x)$ and $B(x)$ defined as: $$
 A(x)=\sum_{k=0}^{[ \frac{N}{2}]}C_{2k}^{N}x^{k},
$$
\begin{equation}
B(x)=\sum_{k=0}^{[ \frac{N-1}{2}]}C_{2k+1}^{N}x^{k}.
\end{equation}
In deriving the of above formula we have used the following
identities:
$$\cos(N\arctan i\sqrt{x})=\frac{A(x)}{(1-x)^{\frac{N}{2}}},\quad
$$
\begin{equation}
\sin(N\arctan i
\sqrt{x})=i\sqrt{x}\frac{B(x)}{(1-x)^{\frac{N}{2}}},\quad
\end{equation}
 inserting the results $(7-3)$ in $(7-2)$, we get:
$$\frac{1+\mbox{a}^{2}y^2}{1+\beta y^2}=\frac{1+\mbox{a}^{2}y^2}{\left(
\frac{B( \frac{1}{\beta})}{\mbox{a} A( \frac{1}{\beta})}+\beta(
\frac{\mbox{a} A( \frac{1}{\beta})}{B(
\frac{1}{\beta})})y^2\right)}.\quad$$
 Hence to get the final result we have to choose the parameter
 $\mbox{a}$ as:
$$\mbox{a} =\frac{B( \frac{1}{\beta})}{A(\frac{1}{\beta})}\quad.$$
 \nonumber
\section{Appendix B}\setcounter{equation}{0}
Here we derive  the invariant measure  of chaotic maps by using
Shure's invariant polynomials in a way which is different form
that of Appendix A. In order to make the paper more readable we
consider only $N=4$ case, i.e, the map,.
\begin{equation}
 y=\frac{1}{\mbox{a}}\tan(4\arctan x)
\end{equation}
 which can be written as:
\begin{equation}
\mbox{a} y=\frac{4x(1-x^{2})}{1-6x^{2}+x^{4}}
\end{equation}
or
\begin{equation}
x^{4}+\frac{4x^{3}}{\mbox{a} y}-6x^{2}-\frac{4x}{\mbox{a} y}+1=0.
\end{equation}
The needed Shure's invariant polynomials of  variables
$x_1,x_2,...,x_4$ are defined as:
$$
S_1=x_1+x_2+x_3+x_4,$$$$
S_{11}=x_1x_2+x_1x_3+x_1x_4+x_2x_3+x_2x_4+x_3x_4,$$$$
S_{111}=x_1x_2x_3+x_1x_2x_4+x_1x_3x_4+x_2x_3x_4,$$\begin{equation}
S_{1111}=x_1x_2x_3x_4 \end{equation}and
$$S_2=x_1^2+x_2^2+x_3^2+x_4^2,$$
$$S_{22}=x_1^2x_2^2+x_1^2x_3^2+x_1^2x_4^2+x_2^2x_3^2+x_2^2x_4^2+x_3^2x_4^2,$$
$$S_{222}=x_1^2x_2^2x_3^2+x_1^2x_2^2x_4^2+x_1^2x_3^2x_4^2+x_2^2x_3^2x_4^2,$$
\begin{equation}
S_{1111}=x_1^2x_2^2x_3^2x_4^2.
\end{equation}
The second set of Shur's invariant polynomials can be expressed
in terms of the first one as:
$$S_2=S_1^2-2S_{11},$$
$$S_{22}=S_{11}^2-2S_1S_{111}+2S_{1111},$$
$$S_{222}=S_{111}^2-2S_{11}S_{1111},$$
\begin{equation}
S_{2222}=S_{1111}^2.
\end{equation}
Considering the variables $x_1, x_2, x_3$ and $ x_4$ as roots of
Equation (8-3), one can obtain the first set of Shur's invariant
polynomials  given in (8-4) as:
\begin{equation}
S_{1}=-\frac{4}{\mbox{a}y},\quad\quad \quad S_{11} =-6, \quad\quad
\quad  S_{111}=\frac{4}{\mbox{a} y},\quad\quad \quad
S_{1111}=1\end{equation} and using Equation (8-6), we have:
\begin{equation}
S_{2}=\frac{16}{(\mbox{a} y)^{2}}+12,\quad
S_{22}=\frac{32}{(\mbox{a} y)^{2}}+38,\quad
S_{222}=\frac{16}{(\mbox{a} y)^{2}}+12, \quad S_{2222}=1.
\end{equation}
Again, writing the  PF equation of map (8-1) and  assuming that
its invariant measure  is of the form (8-3), we have:
\begin{equation}
\mu(y)=\frac{\mbox{a}}{4}\times(\frac{1}{1+\mbox{a}^{2}y^{2}})\sum_{k=1}^{4}\frac{1+x_{k}^{2}}{1+\beta
x_{k}^{2} }.
\end{equation}
The summation on the right-hand side can be written as :
\begin{equation}
\sum_{k=1}^{4}\frac{1+x_{k}^{2}}{1+\beta
x_{k}^{2}}=\frac{(1+x_{1}^{2})(1+\beta x_{2}^{2})(1+\beta
x_{3}^{2})(1+\beta x_{4}^{2})}{\prod_{k=1}^{4}(1+\beta
x_{k}^{2})},
\end{equation}
where using Equation (8-8), the numerator of above fraction
becomes:
\begin{equation}
4+(3\beta+1)S_2+(2\beta^{2}+2\beta)S_{22}+(\beta^{3}+3\beta^{2})S_{222}+4\beta^{3}S_{2222}
= 16(1+\beta)(1+6\beta+\beta^{2}).\end{equation}
 Also using again
(8-8), for its denominator \begin{equation} 1+\beta
S_{2}+\beta^{2} S_{22}+\beta^{3} S_{222}+\beta^{4} S_{2222},
\end{equation} we get:
\begin{equation}
\frac{16}{\mbox{a}^{2}y^{2}}\beta(1+\beta^{2})+(\beta^{2}+6\beta+1)^{2}.
\end{equation}
Using the results that obtained above, the invariant measure takes
the following form :
\begin{equation}
\mu(y)=\frac{4\alpha(1+\beta)(1+6\beta+\beta^{2})}{\alpha^{2}y^{2}(\beta^{2}+6\beta+1)^{2}+16\beta(\beta+1)^{2}}
\end{equation}
which should be equal to $\frac{1}{1+\beta y^{2}}$, where this is
possible only for the following choice of $\mbox{a}$
\begin{equation}
\mbox{a}=\frac{4\beta+4\beta^{2}}{1+6\beta+\beta^{2}}
\end{equation}

\newpage
Figures captions\\
Fig. 1. The schematic diagram of forward and backward branching of
rational order chaotic map for describing simultaneous  production
and consumption of entropy.
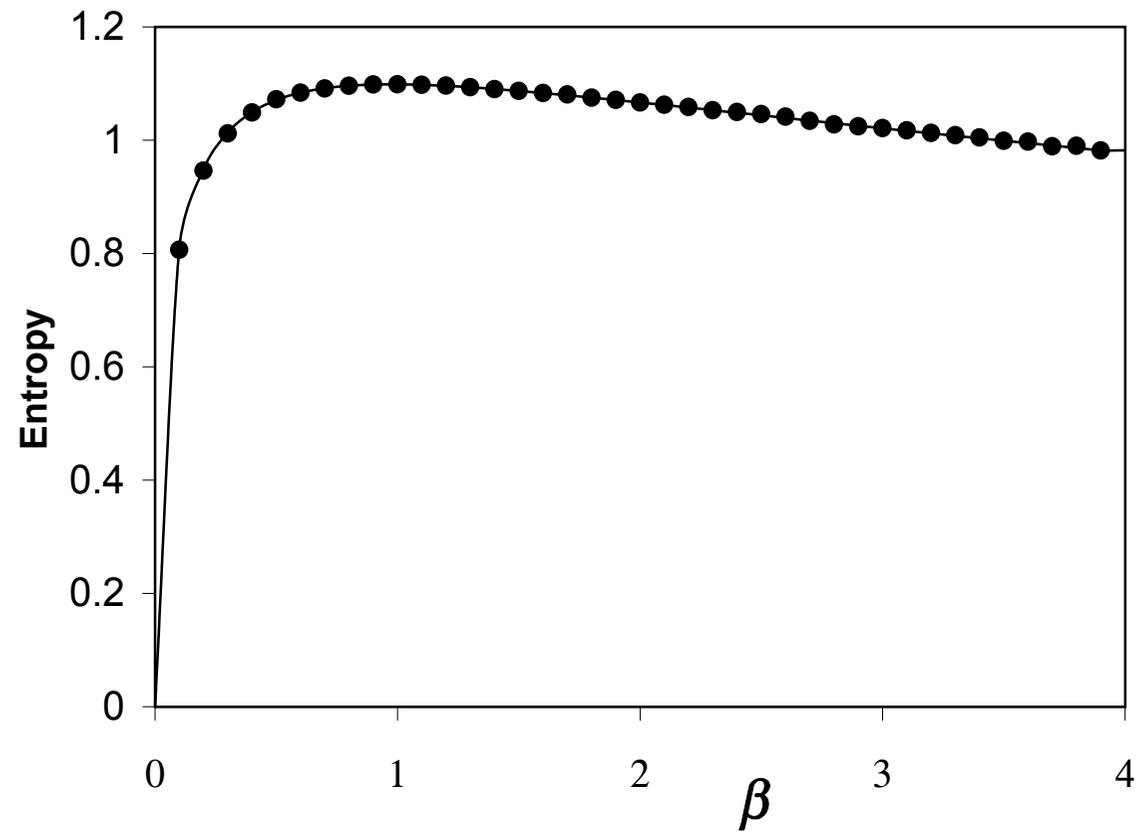
\\Fig. 2. Shows Lyapunov exponent (solid curve) and KS entropy ($\circ$)
versus the control parameter $\beta$, as this figure shows, there
is a maximum  at $\beta$ which corresponds to
$\mbox{a}_{1}=\mbox{a}_{2}=1$.
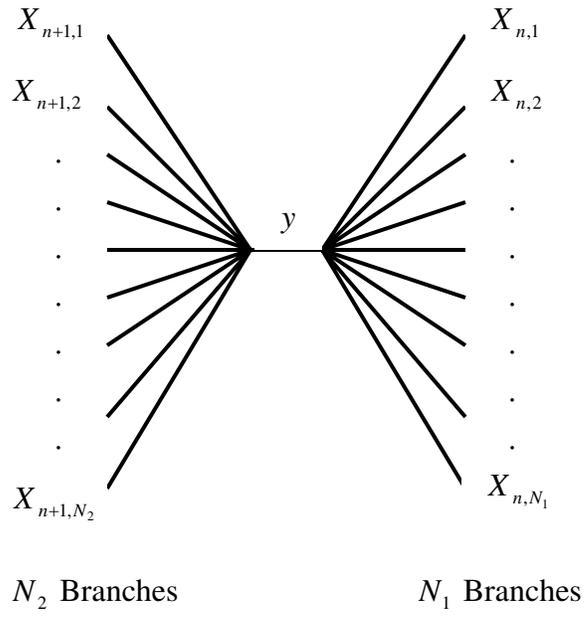
\end{document}